\documentclass[submission, Phys]{SciPost}
\usepackage{graphicx}
\usepackage{amsmath}
\usepackage{amsfonts}
\usepackage{amssymb}
\usepackage{soul}
\usepackage{dsfont}
\usepackage{hyperref}
\usepackage{amstext}

\begin{document}

\begin{center}{\Large \textbf{
Time evolution of interacting bosons through squeezing Hamiltonians
}}\end{center}

\begin{center}
Chester Moore\textsuperscript{1},
David Edward Bruschi\textsuperscript{1*}
\end{center}

\begin{center}
{\bf 1} York Centre for Quantum Technologies, Department of Physics, University of York, Heslington, YO10 5DD York, UK
\\
* david.edward.bruschi@gmail.com
\end{center}

\begin{center}
\today
\end{center}


\section*{Abstract}
{\bf
We study the full time evolution of one- and two-mode bosonic quantum systems that interact through single- and two-mode squeezing Hamiltonians. We establish that the single- and two-mode cases are formally equivalent, leading to the same differential equations encoding the full time evolution.
These differential equations can be easily employed in any application.
We analytically predict a dramatic transition in the population of the modes when the coupling takes a specific critical value, leading to exponential growth of the excitation population. We discuss the validity, scope and generality of our results.
}

\vspace{10pt}
\noindent\rule{\textwidth}{1pt}
\tableofcontents\thispagestyle{fancy}
\noindent\rule{\textwidth}{1pt}
\vspace{10pt}

\section{Introduction}\label{intro}

In this work we investigate the equivalence between time evolution induced by single- and two- mode squeezing.
We employ recently developed techniques to obtain differential equations that govern the full time evolution of the mode operators, which are decoupled, linear and of second order \cite{Bruschi:Lee:2013}. This allows for immediate numerical integration for any specific scenario of interest.
We apply our results to study a variety of parameter regimes, and periodic drives, which lead to exponential population growth.

Our results can be applied to setups with harmonic oscillators, as well to setups that contain modes of a quantum field. Applications of these techniques range from the theory of parametric down-conversion \cite{Rubin:Klyshko:1994}, models of light coupled to nano-mechanical resonators \cite{Aspelmeyer:Kippenberg:2014} and quantum optics \cite{Walborn:Monken:2010} to  single modes of a bosonic field interacting with a large ensemble of two-level systems \cite{Emary:Brandes:2003}, quantum field theory \cite{Birrell:Davies:1984} and cosmology \cite{Arnowitt:Deser:2008}. 

This work is organised as follows. In \autoref{tools} we introduce the necessary tools to be used in this work. In \autoref{time:evolution:main:solution}  we present the analytical results of the time evolution. In \autoref{applications} we discuss applications of our techniques to cases that have been studied or are of interest. Finally, in \autoref{conclusions} we discuss the outlook and possible future directions of this work.

Our conventions are the following: the symbol Tp stands for transposition and, in places, we abbreviate $\cosh$ by $ch$, $\sinh$ by $sh$ and $\tanh$ by $th$ for a better reproduction of results when convenient. We work in the Heisenberg picture.

\section{Tools}\label{tools}
Bosons can be used to model a wide variety of physical systems, from the electromagnetic field propagating in superconducting circuits \cite{Xiang:Ashhab:2013,Wilson:Johansson:2011} and light coupled to nano-mechanical resonators \cite{Aspelmeyer:Kippenberg:2014} to phonons in a Bose-Einstein Condensate \cite{Dalfovo:Giorgini:1999,Carusotto:Ciuti:2013} and radiation emitted by black holes \cite{Hawking:1974}. 
For the sake of clarity and simplicity, in this work we choose to focus our model on harmonic oscillators rather than quantum fields, which have risen as core elements within recent advances in relativistic and quantum information \cite{Alsing:Fuentes:2012}. It is important to note, however, that our techniques and results apply \textit{directly} and in a straightforward fashion to coupled quantum fields, as has been argued before \cite{Bruschi:Lee:2013} (see \cite{Brown:MartinMartinez:2013} for connected results).

\subsection{One and two-mode quadratic Hamiltonians}\label{one:two:mode:quadratic:hamiltonians}

\subsubsection{One-mode quadratic Hamiltonians}\label{tools:qft:single}
We start with one harmonic oscillator, or bosonic mode. We characterise it with (possibly time dependent) frequency $\omega_+$ and annihilation and creation operators $\hat{a},\hat{a}^{\dag}$ that satisfy the canonical commutation relations $[\hat{a},\hat{a}^{\dag}]=1$. The most general time-dependent quadratic Hamiltonian $\mathcal{H}_1$ for one mode is
\begin{align}\label{general:hamiltonian:one:mode}
\hat{\mathcal{H}}_1=\hbar\,\omega_+(t)\,\hat{N}_++2\,\hbar\,\,g_+(t)\,\hat{G}_++2\,\hbar\,g_-(t)\,\hat{G}_-,
\end{align}
where $g_{\pm}$ are time dependent driving functions of dimension frequency and we have defined the basic operator algebra elements $\hat{N}_+:=\frac{1}{2}\,\hat{a}^{\dag}\hat{a}$, $\hat{G}_+:=\frac{1}{4}[\hat{a}^{\dag\,2}+\hat{a}^2]$ and $\hat{G}_-:=i\,(\hat{a}^{\dag\,2}-\hat{a}^2)$ for notational convenience. Notice that $\omega_+=2\,\omega_{\textrm{a}}$ for one mode.

We note here that this Hamiltonian, known in quantum optics as the single mode squeezing Hamiltonian \cite{Mandel:Wolf:1995}, appears, for example, in degenerate parametric down conversion \cite{Rubin:Klyshko:1994}.

We choose to introduce the dimensionless time $\eta:=\frac{\omega_{\textrm{c}}}{2}\,t$, where $\omega_{\textrm{c}}$ is a convenient normalisation frequency whose choice will depend on the situation at hand, and to normalise the Hamiltonian (\ref{general:hamiltonian:one:mode}) by $\hbar$, which allows us to obtain the dimensionless Hamiltonian $H_1(t)$, which reads
\begin{align}\label{general:hamiltonian:one:mode:dimensionless}
\hat{H}_1(\eta)=&4\,\left[\Omega_+(\eta)\,\hat{N}_++h_+(\eta)\,\hat{G}_++h_-(\eta)\,\hat{G}_-\right],
\end{align}
where we have introduced the dimensionless functions $\Omega_+:=\omega_+/\omega_{\textrm{c}}$ and $h_{\pm}(t):=g_\pm(t)/\omega_{\textrm{c}}$.
This choice will prove extremely convenient for our following work.

Notice that we have allowed $\Omega_+$ to be a function of the dimensionless time $\eta$.

\subsubsection{Two-mode quadratic Hamiltonians}\label{tools:qft:twomode}
Let us consider here two harmonic oscillators, with frequencies $\omega_{\textrm{a}}$ and $\omega_{\textrm{b}}$ respectively, and with annihilation and creation operators $\hat{a},\hat{a}^{\dag}$ and $\hat{b},\hat{b}^{\dag}$ which satisfy the canonical commutation relations $[\hat{a},\hat{a}^{\dag}]=[\hat{b},\hat{b}^{\dag}]=1$, while all others vanish. The most general quadratic time-dependent Hamiltonian $\hat{\mathcal{H}}_2(t)$ of these two modes contains $10$ independent elements, i.e., $10$ independent quadratic Hermitian operators with appropriate, perhaps time-dependent, coefficients \cite{Bruschi:Lee:2013}.
Among all possible interaction Hamiltonians $\hat{\mathcal{H}}_2(t)$, we focus here on a simple yet interesting quadratic time-dependent interacting Hamiltonian $\hat{\mathcal{H}}_2(t)$ of the form
\begin{align}\label{general:interaction:hamiltonian}
\hat{\mathcal{H}}_2(t)=&\hbar\,\omega_{\textrm{a}}\,\hat{a}^{\dag}\,\hat{a}+\hbar\,\omega_{\textrm{b}}\,\hat{b}^{\dag}\,\hat{b}+\hbar\,g_+(t)\,\hat{G}_++\hbar\,g_-(t)\,\hat{G}_-,
\end{align}
where $g_{\pm}$ are time dependent driving functions of dimension frequency and we have defined the basic operator algebra elements $\hat{G}_+:=\frac{1}{2}[\hat{a}^{\dag}\hat{b}^{\dag}+\hat{a}\,\hat{b}]$ and $\hat{G}_-:=\frac{i}{2}\,[\hat{a}^{\dag}\hat{b}^{\dag}-\hat{a}\,\hat{b}]$ for notational convenience.
This Hamiltonian, known in quantum optics as the two mode squeezing Hamiltonian \cite{Mandel:Wolf:1995}, appears, for example, in parametric down conversion \cite{Rubin:Klyshko:1994}.

We now introduce the operators $\hat{N}_{\pm}:=\frac{1}{2}\,[\hat{a}^{\dag}\,\hat{a}\pm\hat{b}^{\dag}\,\hat{b}]$ and note that $\hat{N}_-$ commutes simultaneously with $\hat{N}_+$, $\hat{G}_+$ and $\hat{G}_-$. Therefore, the number difference of excitations is a constant of motion. This is a well known result and is a consequence of the fact that the two mode squeezing operation always creates (or annihilates) pairs of excitations coherently \cite{}. This interaction can be engineered in the laboratory by employing nonlinear crystals \cite{Mandel:Wolf:1995}.
We can rewrite the Hamiltonian \eqref{general:interaction:hamiltonian} as
{\small
\begin{align}\label{general:interaction:hamiltonian:intermidiate}
\hat{\mathcal{H}}_2(t)=&\hbar\,\omega_+\,\hat{N}_++\hbar\,\omega_-\,\hat{N}_-+\hbar\,g_+(t)\,\hat{G}_++\hbar\,g_-(t)\,\hat{G}_-,
\end{align}
}
where we have introduced the frequencies $\omega_{\pm}:=\omega_{\textrm{a}}\pm\omega_{\textrm{b}}$. 

We now choose to introduce the dimensionless time $\eta:=\frac{1}{4}\,g_{\textrm{c}}\,t$, where we have introduced an appropriate critical coupling $g_c$. This critical value of the coupling plays an important role in interacting systems within quantum optics \cite{Emary:Brandes:2003,Aspelmeyer:Kippenberg:2014}. 
The interaction Hamiltonian \eqref{general:interaction:hamiltonian:intermidiate}, normalised by $\hbar$, reduces to
\begin{align}\label{first:dimensionless:hamiltonian}
\hat{H}_2(\eta)=&4\,\left[\Omega_+\,\hat{N}_++h_+(\eta)\,\hat{G}_++h_-(\eta)\,\hat{G}_-+\Omega_-\,\,\hat{N}_-\right].
\end{align}
Here have introduced the dimensionless frequencies $\Omega_\pm:=\omega_\pm/g_{\textrm{c}}$ and we have defined $h_\pm:=g_\pm/g_{\textrm{c}}$.

From now on, since $\hat{N}_-$ is a constant of motion, we will drop it from any consideration about time evolution of the system.

\subsection{Symplectic geometry and Covariance Matrix formalism}\label{tools:cmf}
In quantum mechanics, the initial state $\rho_\mathrm{i}$ of a system of $N$ bosonic modes with operators $\{a_n,a^{\dag}_n\}$ evolves to a final state $\rho_\mathrm{f}$ through the standard Heisenberg equation $\rho_\mathrm{f}=U^{\dag}\,\rho_\mathrm{i}\,U$, where $U$ implements the transformation of interest, such as time evolution. If the state $\rho$ is Gaussian and the Hamiltonian $H$ is quadratic in the operators, it is convenient to introduce the vector $\mathbb{X}=(a_1,\ldots,a_N,a^{\dag}_1,\ldots,a^{\dag}_N)^{Tp}$, the vector of first moments $d:=\langle\mathbb{X}\rangle$ and the covariance matrix $\boldsymbol{\sigma}$ defined by $\sigma_{nm}:=\langle\{\mathbb{X}_n,\mathbb{X}^{\dag}_m\}\rangle-2\langle\mathbb{X}_n\rangle\langle\mathbb{X}_m^{\dag}\rangle$, where $\{\cdot,\cdot\}$ stands for anticommutator and all expectation values of an operator $\mathcal{A}$ are defined by $\langle \mathcal{A}\rangle:=\text{Tr}(\mathcal{A}\,\rho)$. In this language, the canonical commutation relations read $[\mathbb{X}_n,\mathbb{X}_m^{\dag}]=i\,\Omega_{nm}$, where the $2N\times2N$ matrix $\boldsymbol{\Omega}$ is known as the symplectic form \cite{Adesso:Ragy:2014}. We then notice that, while arbitrary states of bosonic modes are, in general,  characterised by an infinite amount of degrees of freedom, a Gaussian state is uniquely determined by its first and second moments, $d_n$ and $\sigma_{nm}$ respectively \cite{Adesso:Ragy:2014}. Furthermore, quadratic (i.e., linear) unitary transformations, such as Bogoliubov transformations \cite{Birrell:Davies:1984}, preserve the Gaussian character of the Gaussian state and can always be represented by a $2N\times2N$ symplectic matrix $\boldsymbol{S}$ that preserves the symplectic form, i.e., $\boldsymbol{S}^{\dag}\,\boldsymbol{\Omega}\,\boldsymbol{S}=\boldsymbol{\Omega}$. All of this can be used to show that the Heisenberg equation can be translated in this language to the simple equation $\boldsymbol{\sigma}_\mathrm{f}=\boldsymbol{S}^{\dag}\,\boldsymbol{\sigma}_\mathrm{i}\,\boldsymbol{S}$, which shifts the problem of usually untreatable operator algebra to simple $2N\times2N$ matrix multiplication.  
In addition, Williamson's theorem guarantees that any $2N\times2N$ hermitian matrix, such as the covariance matrix $\boldsymbol{\sigma}$, can be decomposed as $\boldsymbol{\sigma}=\boldsymbol{S}^{\dag}\,\boldsymbol{\nu}_{\oplus}\,\boldsymbol{S}$, where $\boldsymbol{S}$ is an appropriate symplectic matrix, the diagonal matrix $\boldsymbol{\nu}_{\oplus}=\textrm{diag}(\nu_1,\dots,\nu_N,\nu_1,\dots,\nu_N)$ is known as the Williamson form of the state and $\nu_n:=\coth(\frac{2\,\hbar\,\omega_n}{k_B\,T})\geq1$ are the symplectic eigenvalues of the state \cite{Williamson:1936}.

Williamson's form $\boldsymbol{\nu}_{\oplus}$ contains information about the local and global mixedness of the state of the system \cite{Adesso:Ragy:2014}. The state is pure when $\text{det}(\boldsymbol{\sigma})=\text{det}(\boldsymbol{\nu}_{\oplus})=\prod_n \nu_n=1$ and is mixed otherwise. As an example, the thermal state $\boldsymbol{\sigma}_{th}$ of a $N$-mode bosonic system is simply given by its Williamson form, i.e., $\boldsymbol{\sigma}_{th}=\boldsymbol{\nu}_{\oplus}$.

Finally, in the context of symplectic geometry, one has that a quadratic Hamiltonian $H(t)$ can be always represented by the symplectic matrix 
\begin{equation}
\boldsymbol{S}=\overset{\leftarrow}{\mathcal{T}}\,e^{\boldsymbol{\Omega}\,\int_0^{t} dt'\,\boldsymbol{H}(t')},
\end{equation}
where the matrix $\boldsymbol{H}$ can be obtained by $H(\tau)=\mathbb{X}^{\dag}\,\boldsymbol{H}\,\mathbb{X}$.

\subsection{Sub-algebra of the squeezing Hamiltonian}\label{sub:algebra:techniques}
Here we discuss some properties of the sub-algebra generated by $\hat{N}_+$, $\hat{G}_+$ and $\hat{G_-}$ for both single- and two-mode Hamiltonians.

We note that, for both single- and two-mode cases, the commutation of the operators reads
{\small
\begin{align}
\left[\hat{G}_+,\hat{N}_+\right]=i\,\hat{G}_-,\,\,\,\left[\hat{N}_+,\hat{G}_-\right]=i\,\hat{G}_+,\,\,\,\left[\hat{G}_+,\hat{G}_-\right]=i\,\hat{N}_+,
\end{align}
}
a part from an additional constant to the commutator $[\hat{G}_+,\hat{G}_-]$ which reads $i/2$ for the single-mode case and $i$, for the two-mode case. This constant is irrelevant for all purposes of interest to us since contributes to the time evolution through a phase, which has no physical relevance.

The fact that the algebra of the two systems is identical implies that the two Lie groups are homomorphic \cite{Blanes:Casas:2008}. We can therefore forget about which system we are investigating and treat them both at the same time. Interestingly, we note that this algebra is the same that of the Pauli matrices, i.e, the angular momentum algebra.

To see this explicitly we note that the matrices involved in the following, which come from the representation of the generators in matrix form, will be $\boldsymbol{\Omega}$, $\textrm{adg}_\mathds{1}$ and $\textrm{adg}_\mathds{1}\,\boldsymbol{\Omega}$, where $\textrm{adg}_\mathds{1}$ is the anti-diagonal identity. Note that, in the one mode case we have $\sigma_{\textrm{x}}=\textrm{adg}_\mathds{1}$, $\sigma_{\textrm{y}}=-\textrm{adg}_\mathds{1}\,\boldsymbol{\Omega}$ and $\sigma_{\textrm{z}}=i\,\boldsymbol{\Omega}$.

In \autoref{appendix:stability:vacuum} we discuss issues relative to the stability of the vacuum, i.e., the existence of negative eigenvalues.

\section{Time evolution of the system: main solution}\label{time:evolution:main:solution}
Our main goal is to find an explicit expression for the symplectic representation of the operator
\begin{align}
U(\eta)=\overset{\leftarrow}{\mathcal{T}}\,\exp\left[-i\,\int_0^\eta\,d\eta'\,\hat{H}(\eta')\right].
\end{align}
In this section we will avoid printing the explicit expressions of all matrices in order to streamline the discussion. All missing expressions can be found in \autoref{appendix:first:second}.

We start by using the techniques introduced above and we map the problem to the symplectic domain, therefore constructing the symplectic matrix $\boldsymbol{S}(\eta)$ defined by
\begin{align}
\boldsymbol{S}(\eta)=\overset{\leftarrow}{\mathcal{T}}\,\exp\left[\boldsymbol{\Omega}\,\int_0^\eta\,d\eta'\,\boldsymbol{H}(\eta')\right].
\end{align}
The Hamiltonian matrix $\boldsymbol{H}$ can be diagonalised by a time-independent orthogonal matrix when the coupling is real. This is an important feature of the Hamiltonian matrix which we now proceed to exploit. The coupling $h(\eta)$ can be written as $|h(\eta)|\,\exp[i\,\phi(\eta)]$. We then use the results of \autoref{appendix:first} to obtain
\begin{align}
\boldsymbol{S}(\eta)=&\overset{\leftarrow}{\mathcal{T}}\,\exp\left[\boldsymbol{\Omega}\,\int_0^\eta\,d\eta'\,\boldsymbol{H}(\eta')\right]\nonumber\\
=&\overset{\leftarrow}{\mathcal{T}}\,\exp\left[\boldsymbol{\Omega}\,\int_0^\eta\,d\eta'\,\left(\boldsymbol{H}+\boldsymbol{H}_\phi-\boldsymbol{H}_\phi\right)\right]\nonumber\\
=&\boldsymbol{S}_\phi(\eta)\,\overset{\leftarrow}{\mathcal{T}}\,\exp\left[\boldsymbol{\Omega}\,\int_0^\eta\,d\eta'\,\boldsymbol{S}^\dag_\phi(\eta')\,\left(\boldsymbol{H}-\boldsymbol{H}_\phi\right)\,\boldsymbol{S}_\phi(\eta')\right]\nonumber\\
=&\boldsymbol{S}_\phi(\eta)\,\overset{\leftarrow}{\mathcal{T}}\,\exp\left[\boldsymbol{\Omega}\,\int_0^\eta\,d\eta'\,\boldsymbol{H}_{\textrm{r}}(\eta')\right],
\end{align}
where we have introduced the diagonal Hamiltonian $\boldsymbol{H}_\phi=\omega_\phi\,\mathds{1}$ and $\omega_\phi=\frac{d}{d\eta}\phi(\eta)$.

The matrix $\boldsymbol{H}_{\textrm{r}}$ has been defined as $\boldsymbol{H}_{\textrm{r}}:=\boldsymbol{S}^\dag_\phi(\eta)\,\left(\boldsymbol{H}(\eta)-\boldsymbol{H}_\phi\right)\,\boldsymbol{S}_\phi(\eta)$, which has the simple expression $\boldsymbol{H}_{\textrm{r}}=\Omega\,\mathds{1}+h(\eta)\,\textrm{adg}_\mathds{1}$. The dimensionless frequency parameter $\Omega$ is $\Omega:=\Omega_+-\Omega_\phi$, while $\Omega_\phi$ is the dimensionless normalised frequency $\omega_\phi$ normalised by the critical value. The matrix $\boldsymbol{H}_{\textrm{r}}$ can be put in diagonal form by a time-independent, orthogonal matrix $\boldsymbol{M}$, where $\boldsymbol{M}\,\boldsymbol{M}^{\textrm{Tp}}=\mathds{1}$. We have $\boldsymbol{M}^{\textrm{Tp}}\,\boldsymbol{H}_{\textrm{r}}\,\boldsymbol{M}=\boldsymbol{D}_{\textrm{r}}$, where $\boldsymbol{D}_{\textrm{r}}$ is diagonal. In particuar, it can be easily checked that it reads $\boldsymbol{D}_{\textrm{r}}=\Omega\,\mathds{1}+i\,h(\eta)\,\boldsymbol{\Omega}$.

Therefore, we have 
\begin{align}\label{}
\boldsymbol{S}(\eta)=&\boldsymbol{S}_\phi(\eta)\,\overset{\leftarrow}{\mathcal{T}}\,\exp\left[\boldsymbol{\Omega}\,\int_0^\eta\,d\eta'\,\boldsymbol{H}_{\textrm{r}}(\eta')\right]\nonumber\\
=&\boldsymbol{S}_\phi(\eta)\,\overset{\leftarrow}{\mathcal{T}}\,\exp\left[\boldsymbol{\Omega}\,\int_0^\eta\,d\eta'\,\boldsymbol{M}\,\boldsymbol{D}_{\textrm{r}}(\eta')\,\boldsymbol{M}^{\textrm{Tp}}\right]\nonumber\\
=&\boldsymbol{S}_\phi(\eta)\,\overset{\leftarrow}{\mathcal{T}}\,\exp\left[\boldsymbol{M}\,\boldsymbol{K}\,\int_0^\eta\,d\eta'\,\boldsymbol{D}_{\textrm{r}}(\eta')\,\boldsymbol{M}^{\textrm{Tp}}\right]\nonumber\\
=&\boldsymbol{S}_\phi(\eta)\,\boldsymbol{M}\,\overset{\leftarrow}{\mathcal{T}}\,\exp\left[\boldsymbol{K}\,\int_0^\eta\,d\eta'\,\boldsymbol{D}_{\textrm{r}}(\eta')\right]\,\boldsymbol{M}^{\textrm{Tp}},
\end{align}
where $\boldsymbol{K}:=\boldsymbol{M}^{\textrm{Tp}}\,\boldsymbol{\Omega}\,\boldsymbol{M}=i\,\textrm{adg}_\mathds{1}$ is anti-diagonal and we could move the orthogonal matrix $\boldsymbol{M}$ out of the integral, and therefore out of the time-ordered exponential, because it is time independent.

Let us introduce 
\begin{align}\label{SAD:definition}
\boldsymbol{S}_{\textrm{AD}}(\eta):=\overset{\leftarrow}{\mathcal{T}}\,\exp\left[\int_0^\eta\,d\eta'\,\boldsymbol{K}\,\boldsymbol{D}_{\textrm{r}}(\eta')\right].
\end{align}
The fact that $\boldsymbol{K}\,\boldsymbol{D}_{\textrm{r}}(\eta)$ is anti-diagonal allows us to write
\begin{align}\label{auxiliary:SAD}
\boldsymbol{S}_{\textrm{AD}}(\eta):=\boldsymbol{P}+\int_0^\eta d\eta'\,\boldsymbol{K}\,\boldsymbol{D}_{\textrm{r}}\,\boldsymbol{P},
\end{align}
where the \emph{diagonal} matrix $\boldsymbol{P}$ is our new unknown. The formal expression for  $\boldsymbol{P}$ is discussed in \autoref{Appendix:p:matrix}.

We use the fact that $\dot{\boldsymbol{S}}_{\textrm{AD}}(\eta)=\boldsymbol{K}\,\boldsymbol{D}_{\textrm{r}}\,\boldsymbol{S}_{\textrm{AD}}(\eta)$ to find the equation
\begin{align}
\boldsymbol{K}\,\boldsymbol{D}_{\textrm{r}}\,\int_0^\eta d\eta'\,\boldsymbol{K}\,\boldsymbol{D}_{\textrm{r}}\,\boldsymbol{P}=\dot{\boldsymbol{P}}.
\end{align}
The dot is a short notation for derivative with respect to time.

The matrix $\boldsymbol{K}\,\boldsymbol{D}_{\textrm{r}}$ is invertible\footnote{This is true as long as $\Omega\neq h(\eta)$. We will see that one of the analytical solutions considers this case separately.}, therefore we can employ some algebra and obtain
\begin{align}\label{main:differential:equation:compact:matrix:form}
\ddot{\boldsymbol{P}}-\boldsymbol{K}\,\dot{\boldsymbol{D}}_{\textrm{r}}\,\boldsymbol{D}^{-1}_{\textrm{r}}\,\boldsymbol{K}^\dag\,\dot{\boldsymbol{P}}-(\boldsymbol{K}\,\boldsymbol{D}_{\textrm{r}})^2\,\boldsymbol{P}=0,
\end{align}
which collects our main set of differential equations in a compact form. We have used the fact that $\boldsymbol{K}^\dag=\boldsymbol{K}^{-1}$. As a consistency check we note that both $\boldsymbol{K}\,\dot{\boldsymbol{D}}_{\textrm{r}}\,\boldsymbol{D}^{-1}_{\textrm{r}}\,\boldsymbol{K}^\dag$ and $(\boldsymbol{K}\,\boldsymbol{D}_{\textrm{r}})^2$ are diagonal matrices.

We also note that, since $\boldsymbol{P}$ is diagonal, the differential equations for the elements $P_{nn}$ of the matrix $\boldsymbol{P}$, which are the only non-zero elements, are all decoupled.

The differential equation \eqref{main:differential:equation:compact:matrix:form} needs to be complemented with two initial conditions. The first one simply requires that $\boldsymbol{P}(0)=0$, while the second can be obtained by taking the first derivative of \eqref{auxiliary:SAD}, equating it to the time derivative of the definition of $\boldsymbol{S}_{\textrm{AD}}(\eta)$ and then evaluating at $t=0$. This condition reads $\dot{\boldsymbol{P}}(0)=0$.

We can look at \autoref{appendix:first:second} and at the form of the Hamiltonian matrices \eqref{eqn:hammatrix:single:appendix} and \eqref{eqn:hammatrix:appendix}. 
We note that, whatever the number of modes, we have $(\boldsymbol{K}\,\boldsymbol{D}_{\textrm{r}})^2=-\rho^2\,\mathds{1}$, where $\rho^2:=\Omega^2-h^2(\eta)$ and we have defined $\Omega(\eta):=\rho\,\cosh \chi$ and $h(\eta):=\rho\,\sinh \chi$. The variables $\rho$ and $\chi$ are functions of $\eta$. Furthermore, it is easy to check that
\begin{align}
\boldsymbol{K}\,\dot{\boldsymbol{D}}_{\textrm{r}}\,\boldsymbol{D}^{-1}_{\textrm{r}}\,\boldsymbol{K}^\dag=\frac{\dot{\rho}}{\rho}\,\mathds{1}-i\,\frac{\dot{\Omega}\,h-\Omega\,\dot{h}}{\rho^2}\,\boldsymbol{\Omega}=\frac{\dot{\rho}}{\rho}\,\mathds{1}+i\,\dot{\chi}\,\boldsymbol{\Omega}.
\end{align}
Putting all together, and using some algebra, it is easy to check that main differential equations \eqref{main:differential:equation:compact:matrix:form} reduce to the following two un-couplued, second order, linear differential equations
\begin{align}\label{main:differential:equation:compact:matrix:form:good}
\ddot{p}_1-\left(\frac{\dot{\rho}}{\rho}+\dot{\chi}\right)\,\dot{p}_1+\rho^2\,p_1=&0\nonumber\\
\ddot{p}_2-\left(\frac{\dot{\rho}}{\rho}-\dot{\chi}\right)\,\dot{p}_2+\rho^2\,p_2=&0.
\end{align}
Note also that, for the two mode case, one has $p_{22}=p_{11}=p_1$ and $p_{44}=p_{33}=p_2$. 
The expressions \eqref{main:differential:equation:compact:matrix:form:good} complement and complete the expression
\begin{align}\label{final:symplectic:matrix}
\boldsymbol{S}(\eta)=&\boldsymbol{S}_\phi(\eta)\,\boldsymbol{M}\,\boldsymbol{S}_{\textrm{AD}}(\eta)\,\boldsymbol{M}^{\textrm{Tp}},
\end{align}
and are the only non-analytical features of this work. We can write the expression $\boldsymbol{S}(\eta)=\boldsymbol{S}_\phi(\eta)\,\boldsymbol{S}_{\textrm{sq}}(\eta)$, where the explicit expression for matrix $\boldsymbol{S}_{\textrm{sq}}$ can be found in \autoref{appendix:first:second} for both cases.

We can introduce $p_1(\eta)=p_1(y(\eta))$,  and $p_2(\eta)=p_2(y(\eta))$, where $y(\eta):=\int_0^\eta\,d\eta'\,\rho(\eta')$. Then, the main differential equations \eqref{main:differential:equation:compact:matrix:form:good} take the alternative form
\begin{align}\label{main:differential:equation:compact:matrix:form:alternative}
\ddot{p}_1-\dot{\chi}\,\dot{p}_1+p_1=&0\nonumber\\
\ddot{p}_2+\dot{\chi}\,\dot{p}_2+p_2=&0,
\end{align}
where the derivatives are now with respect to $y$ and we have introdouce the implicit definition $\frac{d}{dy}\chi=\frac{1}{\rho}\,\frac{d}{d\eta}\chi$.

It is easy to show that by introducing $p_{\pm}:=p_1\pm p_2$ and 
\begin{align}
I_\pm:=I_1\pm I_2=\int_0^\eta d\eta'\,\left[p_1\,(\Omega+h(\eta'))\pm p_2\,(\Omega-h(\eta'))\right]
\end{align}
it follows that the symplectic matrix $\boldsymbol{S}(\eta)$ is then defined uniquely by the two Bogoliubov coefficients
\begin{align}
\alpha=&\frac{1}{2}\,e^{-i\,\phi}\,\left[p_++i\,I_+\right]\nonumber\\
\beta=&\frac{1}{2}\,e^{-i\,\phi}\,\left[p_--i\,I_-\right],
\end{align}
which satisfy the Bogoliubov identity $|\alpha^2|-|\beta|^2=1$, which reads
\begin{align}
1=&p_1\,p_2+I_1\,I_2,
\end{align}
while the second identity $\alpha\,\beta^{Tp}-\beta\,\alpha^{Tp}=0$ is automatically satisfied.

As a consistency check, note that when $h=0$ we have that $\boldsymbol{S}_\phi(\eta)=\mathds{1}$, $p_1=p_2=\cos(\Omega\,\eta)$ and therefore from \eqref{orthogonal:matrix:single:appendix} or \eqref{orthogonal:matrix:two:mode:appendix} we find that \eqref{final:symplectic:matrix} reduces to just the free evolution sympletic matrix, as expected.

These expression cannot be simplified further, but we will show it has solutions for situations of interest.

\section{Time evolution of the system: solutions for cases of interest}\label{applications}
We now proceed to show that the main solution \eqref{main:differential:equation:compact:matrix:form:good} has analytical expression for cases of broad interest.

\subsection{Parametric drive}
Here we assume that $h(\eta)=h_0\,\Omega_+\,\cos(2\,\Omega_{\textrm{d}}\,\eta)$, that $\Omega=\Omega_+-h(\eta)$ and that $\Omega_+$ is constant.
It is easy to show that the main differential equations \eqref{main:differential:equation:compact:matrix:form:good} reduce to
\begin{align}\label{main:differential:equation:compact:matrix:form:parametric:drive}
0=&\ddot{p}_1+\Omega_+^2\,\left(1-2\,h_0\,\cos(2\,\Omega_{\textrm{d}}\,\eta)\right)\,p_1\nonumber\\
0=&\ddot{p}_2+4\,\Omega_{\textrm{d}}\,h_0\,\frac{\sin(2\,\Omega_{\textrm{d}}\,\eta)}{1-2\,h_0\,\cos(2\,\Omega_{\textrm{d}}\,\eta)}\,\dot{p}_2+\Omega_+^2\,\left(1-2\,h_0\,\cos(2\,\Omega_{\textrm{d}}\,\eta)\right)\,p_2.
\end{align}
We note that the first equation is the well known Mathieu equation which naturally arises in the context of parametrically driven harmonic oscillators and whose solutions for different parameter regimes are known \cite{NIST:2010}. 

\subsection{Periodic drive}
Let us assume that the coupling is time dependent and that the time dependence is periodic, i.e., the system is driven with a coupling of the form $h(t)=h_0\,e^{-i\,\Omega_{\textrm{d}}\,\eta}$, which oscillates with demensionless drive frequency $\Omega_{\textrm{d}}$ and that $\Omega_+$ is time-independent. In this case $\rho_0^2=\Omega^2-h_0^2$ and is time independent.

Some algebra allows us to solve \eqref{main:differential:equation:compact:matrix:form:good}  and obtain
\begin{align}
\alpha=&e^{-i\,\Omega_{\textrm{d}}\,\eta}\,\left[\cos(\rho_0\,\eta)+i\,\frac{\Omega}{\rho_0}\,\sin(\rho_0\,\eta)\right]\nonumber\\
\beta=&i\,e^{-i\,\Omega_{\textrm{d}}\,\eta}\,\frac{h_0}{\rho_0}\,\sin(\rho_0\,\eta).
\end{align}
As an application, we know that the time evolution of the operator $\hat{a}$ is $\hat{a}=\alpha(\eta)\,\hat{a}+\beta(\eta)\,\hat{a}^\dag$ and therefore we can compute the time-dependent expectation value of the number operator $\langle\hat{a}^\dag\hat{a}\rangle(\eta)$, which reads
\begin{align}
\langle\hat{a}^\dag\hat{a}\rangle(\eta)=&\left(1+2\,|\beta|^2\right)\langle\hat{a}^\dag\hat{a}\rangle+|\beta|^2+\alpha^*\,\beta\,\langle\hat{a}^{\dag2}\rangle+\alpha\,\beta^*\,\langle\hat{a}^2\rangle.
\end{align}
In the present case we have
\begin{align}
\langle\hat{a}^\dag\hat{a}\rangle(\eta)=&\left(1+2\,\frac{h^2_0}{\rho_0^2}\,\sin^2(\rho_0\,\eta)\right)\langle\hat{a}^\dag\hat{a}\rangle+\frac{h^2_0}{\rho_0^2}\,\sin^2(\rho_0\,\eta)+\frac{1}{2}\,i\,\frac{h_0}{\rho_0}\,\sin(2\,\rho_0\,\eta)\,\left(\langle\hat{a}^{\dag2}\rangle-\langle\hat{a}^2\rangle\right)\nonumber\\
&+\frac{h_0\,\Omega}{\rho_0^2}\,\sin^2(\rho_0\,\eta)\,\left(\langle\hat{a}^{\dag2}\rangle+\langle\hat{a}^2\rangle\right).
\end{align}

\subsection{Degenerate coupling}
Let us assume that the the coupling is such that of the form $h(\eta)=\Omega(\eta)$. This case includes the one where the Hamiltonian reduces to a pure $\hat{x}^2$ or $\hat{p}^2$-like expression. 

The main solution \eqref{main:differential:equation:compact:matrix:form:good} cannot be computed in the way that we have presented. Instead, we trace back to the definition \eqref{SAD:definition} of which we reprint here
\begin{align}
\boldsymbol{S}_{\textrm{AD}}(\eta)=\overset{\leftarrow}{\mathcal{T}}\,\exp\left[\int_0^\eta\,d\eta'\,\boldsymbol{K}\,\boldsymbol{D}_{\textrm{r}}(\eta')\right].
\end{align}
We note that $\boldsymbol{K}\,\boldsymbol{D}_{\textrm{r}}$ has the expression $2\,h(\eta)\,\boldsymbol{T}$, which is diagonal and cannot be inverted. Given this expression, and the fact that $\boldsymbol{T}^2=0$, we can easily show that
\begin{align}
\boldsymbol{S}_{\textrm{AD}}(\eta)=\mathds{1}+2\,i\,H(\eta)\,\boldsymbol{T},
\end{align}
where we have defined $H(\eta):=\int_0^\eta\,d\eta'\,h(\eta')$.

This allows us to easily find the explicit expression for $\boldsymbol{S}(\eta)$ in our case. Namely, we have that
\begin{align}
\alpha=&e^{-i\,\phi}\,\left[1+i\,H(\eta)\right]\nonumber\\
\beta=&-e^{-i\,\phi}\,H(\eta).
\end{align}
We can compute again the time evolution of the operator $\hat{a}$, which in the present case reads
\begin{align}
\langle\hat{a}^\dag\hat{a}\rangle(\eta)=&\left(1+2\,H^2(\eta)\right)\langle\hat{a}^\dag\hat{a}\rangle+H^2(\eta)+\left(1-i\,H(\eta)\right)\,H(\eta)\langle\hat{a}^{\dag2}\rangle+\left(1+i\,H(\eta)\right)\,H(\eta)\langle\hat{a}^2\rangle.
\end{align}

\section{Considerations on the results}
Here we address a few important issues that relate to our work.

\subsection{Population ``explosion'' with Periodic drive}
We have found that, for the periodic drive case,
\begin{align}\label{bogo:periodic}
\alpha=&e^{-i\,\Omega_{\textrm{d}}\,\eta}\,\left[\cos(\rho_0\,\eta)+i\,\frac{\Omega}{\rho_0}\,\sin(\rho_0\,\eta)\right]\nonumber\\
\beta=&i\,e^{-i\,\Omega_{\textrm{d}}\,\eta}\,\frac{h_0}{\rho_0}\,\sin(\rho_0\,\eta).
\end{align}
Notice that, if we had $\Omega<h_0$, then we would have that $\rho_0=i\,\sqrt{h_0^2-\Omega^2}=i\,\tilde{\rho}_0$. This would imply that \eqref{bogo:periodic} become
\begin{align}\label{bogo:periodic:explode}
\alpha=&e^{-i\,\Omega_{\textrm{d}}\,\eta}\,\left[\cosh(\tilde{\rho}_0\,\eta)-i\,\frac{\Omega}{\tilde{\rho}_0}\,\sinh(\tilde{\rho}_0\,\eta)\right]\nonumber\\
\beta=&e^{-i\,\Omega_{\textrm{d}}\,\eta}\,\frac{h_0}{\rho_0}\,\sinh(\tilde{\rho}_0\,\eta)
\end{align}
and expectation values such as the mean occupation number $\langle\hat{a}^\dag\hat{a}\rangle(\eta)$ would diverge with time.
This transition point is known in Dicke-like models \cite{Emary:Brandes:2003}

\subsection{Validity and scope of the results}
The results of our work are general, in the sense that they apply to arbitrary quadratic, squeezing-like Hamiltonians of bosonic fields or modes. The results do not depend on the quantisation scheme and can therefore employed also when dealing with relativistic quantum fields in the framework of quantum field theory in curved spacetime \cite{Alsing:Fuentes:2012}. In this context, one needs to pay particular attention and care to the canonical commutation relations, which formally give rise to Dirac-deltas. This issue can be circumvented by using localised quantum fields, such as bosonic fields of light confined in cavities \cite{Alsing:Fuentes:2012}.

Our results are analytical, although the central quantities that appear in the Bogoliubov coefficients have to be obtained by solving a differential equation which generally does not admit an exact solution. This implies that solutions must be found numerically. We stress here, however, that the main differential equations \eqref{main:differential:equation:compact:matrix:form:good}, or any variation that can be obtained by other manipulations, are ordinary, linear, un-coupled and second order. This allows for extremely efficient numerical integration.

\section{Conclusion}\label{conclusions}
In this work we studied the time evolution of coupled one- and two-mode bosonic systems that interact with a time dependent squeezing Hamiltonian. 
We discussed the formal equivalence between these two Hamiltonians and we provided a set of simple uncoupled, second-order differential equations that allow for immediate numerical integration.
We have applied our results to cases of interest, such as periodic drive and parametric drives, and we have discussed the existence of parameter values where there is a dramatic transition in the average population of the modes.
Our results can be used to obtain better understanding in the study of quadratic bosonic systems.

\section*{Acknowledgments}
We acknowledge Chaitanya Joshi, Elinor Irish, Leila Khouri, Antony Lee, Jan Kohlrus, Daniele Faccio, Tim Spiller, Dennis R\"atzel, Sofia Qvafort, Fabienne Schneiter, Ana Luc\'ia B\'aez-Camargo Aguilar and Luis Cortes-Barbado for useful comments and discussions. We extend particular thanks to Andr\'e Xuereb for invaluable help with analysing the differential equations and to Jorma Louko for aid with analysing the solutions. D.E.B. also acknowledges partial support from the COST Action MP1405 QSPACE.

\appendix

\section{Splitting of a time ordered exponential operator}\label{appendix:first}
Here we show that, given time ordered operator 
\begin{align}\label{time:ordered:operator:breakup:start}
\hat{U}(t)=\overset{\leftarrow}{\mathcal{T}}\exp\left[-\frac{i}{\hbar}\int dt'\,\hat{G}(t')\right],
\end{align}
and given \emph{any} decomposition of the Hermitian operator $\hat{G}$ as $\hat{G}=\hat{G}_0+\hat{G}_1$, we can write \eqref{time:ordered:operator:breakup:start} as
\begin{align}\label{time:ordered:operator:breakup:claim}
\hat{U}(t)=\overset{\leftarrow}{\mathcal{T}}\exp\left[-\frac{i}{\hbar}\int dt'\,\hat{G}_0(t')\right]\,\times\overset{\leftarrow}{\mathcal{T}}\exp\left[-\frac{i}{\hbar}\int dt'\,\hat{U}^\dag_0(t')\,\hat{G}_1(t')\,\hat{U}_0(t')\right],
\end{align}
where we have defined 
\begin{align}
\hat{U}_0(t):=\overset{\leftarrow}{\mathcal{T}}\exp\left[-\frac{i}{\hbar}\int dt'\,\hat{G}_0(t')\right].
\end{align}
Notice that the choice of the split of the operator $\hat{G}$ is \emph{arbitrary}, and that the expression\eqref{time:ordered:operator:breakup:claim} is exact.

This is easy to prove. We first take the time derivative of the operator $\hat{U}$ and use the expression \eqref{time:ordered:operator:breakup:start} to find $\frac{d}{dt}\hat{U}(t)=-\frac{i}{\hbar}\,\hat{G}(t)\,\hat{U}(t)$.
We then take the time derivative of the operator $\hat{U}$ and use the expression \eqref{time:ordered:operator:breakup:claim} and we find
\begin{align}
\frac{d}{dt}\hat{U}(t)=&-\frac{i}{\hbar}\,\left[\hat{G}_0(t)\,\hat{U}(t)+\hat{U}_0(t)\,\hat{U}^\dag_0(t)\,\hat{G}_1(t)\,\hat{U}(t)\right]\nonumber\\
=&-\frac{i}{\hbar}\,\left[\hat{G}_0(t)\,\hat{U}(t)+\hat{G}_1(t)\,\hat{U}(t)\right]\nonumber\\
=&-\frac{i}{\hbar}\,\hat{G}(t)\,\hat{U}(t),
\end{align}
which proves that the time derivative of the two expressions is the same.
Given that the solution of two identical first order equations with the same initial conditions (i.e., $\hat{U}(0)=\mathds{1}$) is the same, we have proven that \eqref{time:ordered:operator:breakup:claim} is an alternative expression for \eqref{time:ordered:operator:breakup:start}, which was our claim.

Notice that our relation is valid for time-ordered exponentials of matrices as well. This is not surprising given that matrices are, in the end, a particular representation of linear operators.

Finally, we add a remark. The Hermitian conjugate $\hat{U}^\dag(t)$ of $\hat{U}(t)$ has the expression 
\begin{align}
\hat{U}^\dag(t)=\overset{\rightarrow}{\mathcal{T}}\exp\left[\frac{i}{\hbar}\int dt'\,\hat{G}(t')\right].
\end{align}
Notice that the ordering of the operators, in the expansion, needs to be \emph{reversed}, i.e., ordered from left to right instead of right to left with increasing time. This is symbolized by the reversed arrow.

\section{Definitions of matrix quantities used in this work}\label{appendix:first:second}
Here we list explicit expressions for the one- and two- mode quantities used throughout the paper that have not been listed in the text to avoid cumbersome notation.

The matrix representation of the \emph{full} one- and two-mode Hamiltonians is
\begin{align}
\label{eqn:hammatrix:single:appendix}
\mathbf{H}=
\left(\begin{array}{cc} 
\Omega_+ & h(\eta) \\
h^*(\eta) & \Omega_+
\end{array}\right)
\end{align}
and
\begin{align}
\label{eqn:hammatrix:appendix:preliminary}
\mathbf{H}=
\left(\begin{array}{cccc} 
\Omega_++\Omega_- & 0 & 0 & h(\eta) \\
0 & \Omega_+-\Omega_- & h(\eta) & 0 \\
0 & h^*(\eta) & \Omega_++\Omega_- & 0 \\
h^*(\eta) & 0 & 0 & \Omega_+-\Omega_-
\end{array}\right)
\end{align}
respectively.
For the purpose of the time evolution calculations, we drop the $\hat{N}_-$ term of the Hamiltonian and we are left with
\begin{align}
\label{eqn:hammatrix:appendix}
\mathbf{H}=
\left(\begin{array}{cccc} 
\Omega_+ & 0 & 0 & h(\eta) \\
0 & \Omega_+ & h(\eta) & 0 \\
0 & h^*(\eta) & \Omega_+ & 0 \\
h^*(\eta) & 0 & 0 & \Omega_+
\end{array}\right),
\end{align}
which we will be analysing in the text.

An arbitrary symplectic matrix $\boldsymbol{S}$ has the expression
\begin{align}
\label{arbitrary:symplectic:marix:appendix}
\mathbf{S}=
\left(\begin{array}{cc} 
\boldsymbol{\alpha} & \boldsymbol{\beta} \\
\boldsymbol{\beta}^*  & \boldsymbol{\alpha}^*
\end{array}\right),
\end{align}
where the $N\times N$ matrices $\boldsymbol{\alpha}$ and $\boldsymbol{\beta}$ collect the well-known Bogoliubov coefficients.

The orthogonal rotation matrix $\boldsymbol{M}$ reads 
\begin{align}
\label{orthogonal:matrix:single:appendix}
\boldsymbol{M}=
\frac{1}{\sqrt{2}}
\left(\begin{array}{cc} 
1 & -1 \\
1 & 1
\end{array}\right)
\end{align}
and
\begin{align}
\label{orthogonal:matrix:two:mode:appendix}
\boldsymbol{M}=
\frac{1}{\sqrt{2}}
\left(\begin{array}{cccc} 
1 & 0 & 0 & -1 \\
0 & 1 & -1 & 0 \\
0 & 1 & 1 & 0 \\
1 & 0 & 0 & 1
\end{array}\right).
\end{align}
The matrix $\boldsymbol{P}$ has the general expression
\begin{align}
\label{orthogonal:matrix:single:appendix}
\boldsymbol{P}=
\left(\begin{array}{cc} 
p_{11} & 0 \\
0 & p_{22}
\end{array}\right),
\,\,\,\,\,\,\,\,\,\,\,
\boldsymbol{P}=
\left(\begin{array}{cccc} 
p_{11} & 0 & 0 & 0 \\
0 & p_{22} & 0 & 0 \\
0 & 0 & p_{33} & 0 \\
0 & 0 & 0 & p_{44}
\end{array}\right),
\end{align}
and the differential equations for the two-mode case show us that $p_{22}=p_{11}$ and $p_{44}=p_{33}$.

The matrix $\boldsymbol{S}_{\textrm{sq}}$ reads
\begin{align}
\label{orthogonal:matrix:single:appendix}
\boldsymbol{S}_{\textrm{sq}}=
\frac{1}{2}
\left(\begin{array}{cc} 
p_+-i\,I_+ & p_--i\,I_- \\
p_-+i\,I_- & p_++i\,I_+
\end{array}\right)
\end{align}
and
\begin{align}
\label{orthogonal:matrix:two:mode:appendix}
\boldsymbol{S}_{\textrm{sq}}=
\frac{1}{2}
\left(\begin{array}{cccc} 
p_+-i\,I_+ & 0 & 0 & p_--i\,I_- \\
0 & p_+-i\,I_+ & p_--i\,I_- & 0 \\
0 & p_-+i\,I_- & p_++i\,I_+ & 0 \\
p_-+i\,I_- & 0 & 0 & p_++i\,I_+
\end{array}\right)
\end{align}
for the two cases respectively. 

The degenerate matrix $\boldsymbol{T}$ reads
\begin{align}
\boldsymbol{T}=
\left(\begin{array}{cc} 
0 & 0 \\
1 & 0
\end{array}\right),
\,\,\,\,\,\,\,\,\,\,\,
\boldsymbol{T}=
\left(\begin{array}{cccc} 
0 & 0 & 0 & 0 \\
0 & 0 & 0 & 0 \\
0 & 1 & 0 & 0 \\
1 & 0 & 0 & 0
\end{array}\right)
\end{align}
for the two cases respectively.

\section{Time ordered exponentials}\label{Appendix:p:matrix}
We now look at \eqref{SAD:definition}, which we reprint here
\begin{align}
\boldsymbol{S}_{\textrm{AD}}(\eta):=\overset{\leftarrow}{\mathcal{T}}\,\exp\left[\int_0^\eta\,d\eta'\,\boldsymbol{K}\,\boldsymbol{D}_{\textrm{r}}(\eta')\right].
\end{align}
This has the formal expression
\begin{align}
\boldsymbol{S}_{\textrm{AD}}(\eta)=&\mathds{1}+\int_0^\eta\,d\eta'\,\boldsymbol{K}\,\boldsymbol{D}_{\textrm{r}}(\eta')+\int_0^\eta\,d\eta'\,\boldsymbol{K}\,\boldsymbol{D}_{\textrm{r}}(\eta')\int_0^{\eta'}\,d\eta''\,\boldsymbol{K}\,\boldsymbol{D}_{\textrm{r}}(\eta'')\nonumber\\
&+\int_0^\eta\,d\eta'\,\boldsymbol{K}\,\boldsymbol{D}_{\textrm{r}}(\eta')\int_0^{\eta'}\,d\eta''\,\boldsymbol{K}\,\boldsymbol{D}_{\textrm{r}}(\eta'')\int_0^{\eta''}\,d\eta'''\,\boldsymbol{K}\,\boldsymbol{D}_{\textrm{r}}(\eta''')\nonumber\\
&+\int_0^\eta\,d\eta'\,\boldsymbol{K}\,\boldsymbol{D}_{\textrm{r}}(\eta')\int_0^{\eta'}\,d\eta''\,\boldsymbol{K}\,\boldsymbol{D}_{\textrm{r}}(\eta'')\int_0^{\eta''}\,d\eta'''\,\boldsymbol{K}\,\boldsymbol{D}_{\textrm{r}}(\eta''')\int_0^{\eta'''}\,d\eta''''\,\boldsymbol{K}\,\boldsymbol{D}_{\textrm{r}}(\eta'''')+\ldots
\end{align}
By introducing the matrix $\boldsymbol{P}$ withe expression
\begin{align}
\boldsymbol{P}:=&\mathds{1}+\int_0^\eta\,d\eta'\,\boldsymbol{K}\,\boldsymbol{D}_{\textrm{r}}(\eta')\int_0^{\eta'}\,d\eta''\,\boldsymbol{K}\,\boldsymbol{D}_{\textrm{r}}(\eta'')\nonumber\\
&+\int_0^\eta\,d\eta'\,\boldsymbol{K}\,\boldsymbol{D}_{\textrm{r}}(\eta')\int_0^{\eta'}\,d\eta''\,\boldsymbol{K}\,\boldsymbol{D}_{\textrm{r}}(\eta'')\int_0^{\eta''}\,d\eta'''\,\boldsymbol{K}\,\boldsymbol{D}_{\textrm{r}}(\eta''')\int_0^{\eta'''}\,d\eta''''\,\boldsymbol{K}\,\boldsymbol{D}_{\textrm{r}}(\eta'''')+\ldots
\end{align}
it is easy to see that 
\begin{align}
\boldsymbol{S}_{\textrm{AD}}(\eta)=&\boldsymbol{P}+\int_0^\eta\,d\eta'\,\boldsymbol{K}\,\boldsymbol{D}_{\textrm{r}}(\eta')\,\boldsymbol{P}(\eta'),
\end{align}
which is our claim in the main text.

\section{Stability of the vacuum}\label{appendix:stability:vacuum}
We discuss here another important issue that arises when studying arbitrary Hamiltonians and stability of classical and quantum systems. An arbitrary Hamiltonian is a Hermitian operator with real eigenvalues, however, in order for it to represent a physical process characterised by a spectrum of energies bounded from below (or, equivalently, with a stable vacuum state), the eigenvalues must be positive \cite{Messiah:1961}. It is well known that the presence of one (or more) points where the Hamiltonian ceases to have only positive real eigenvalues is a signature of quantum phase transitions \cite{Emary:Brandes:2003}. Furthermore, the question of the stability of the ground state of bosonic systems with time dependent potentials is of great importance for the understanding of the dynamics of these systems. Conditions on the stability in experimentally meaningful potentials, such as a periodic monochromatic wave, have been found in the literature \cite{Berman:James:2001}.

\subsubsection{Stability of the vacuum: single mode}
Let us look at our Hamiltonian (\ref{first:dimensionless:hamiltonian}). It can be easily put in matrix form, i.e., it is immediate to find the matrix $\boldsymbol{H}$ that represents it from the relation $\mathcal{H}=\mathbb{X}^{\dag}\,\boldsymbol{H}\,\mathbb{X}$. This has the expression \eqref{eqn:hammatrix:single:appendix}, where we have introduced the complex strength $h(\eta):=h_++i\,h_-$.
We compute the eigenvalues $\lambda_{\pm}$ of the matrix \eqref{eqn:hammatrix:single:appendix} which take the expression
\begin{align}\label{two:mode:squeezing:hamiltonian:eigenvalues}
\lambda_{\pm}=\Omega_+\pm|h(\eta)|.
\end{align}
It is immediate to see that $\lambda_+>0$ for any value of the parameters, however, $\lambda_-$ is positive only when the renormalised dimensionless coupling $h$ satisfies $|h|\leq1$, which translates to the well known bound $g(\eta)\leq g_c$ for the dimensional coupling $g(\eta)$. 

We conclude that, also in our case, the Hamiltonian (\ref{first:dimensionless:hamiltonian}) can be used only for couplings that do not exceed the critical value $g_c$.

\subsubsection{Stability of the vacuum: two modes}
We proceed in the same fashion as for one mode. We look at the Hamiltonian \eqref{first:dimensionless:hamiltonian} and put in matrix form. The expression is \eqref{eqn:hammatrix:appendix}, where we have introduced the complex strength $h(\eta):=h_++i\,h_-$.
We compute the eigenvalues $\lambda_{\pm}$ of the matrix \eqref{eqn:hammatrix:appendix} which are doubly degenerate and take the expression
\begin{align}\label{two:mode:squeezing:hamiltonian:eigenvalues}
\lambda_{\pm}=\Omega_+\pm\sqrt{\Omega^2_-+|h(\eta)|^2}.
\end{align}
It is immediate to see that $\lambda_+>0$ for any value of the parameters, however, $\lambda_-$ is positive only when the renormalised dimensionless coupling $h$ satisfies $|h|\leq\sqrt{\Omega_+^2-\Omega_-^2}$.

\bibliography{DetectorsBib}

\begin{thebibliography}{10}
\providecommand{\url}[1]{\texttt{#1}}
\providecommand{\urlprefix}{URL }
\expandafter\ifx\csname urlstyle\endcsname\relax
  \providecommand{\doi}[1]{doi:\discretionary{}{}{}#1}\else
  \providecommand{\doi}{doi:\discretionary{}{}{}\begingroup
  \urlstyle{rm}\Url}\fi
\providecommand{\eprint}[2][]{\url{#2}}

\bibitem{Bruschi:Lee:2013}
D.~E. Bruschi, A.~R. Lee and I.~Fuentes,
\newblock \emph{Time evolution techniques for detectors in relativistic quantum
  information},
\newblock Journal of Physics A: Mathematical and Theoretical \textbf{46}(16),
  165303 (2013).

\bibitem{Rubin:Klyshko:1994}
M.~H. Rubin, D.~N. Klyshko, Y.~H. Shih and A.~V. Sergienko,
\newblock \emph{Theory of two-photon entanglement in type-ii optical parametric
  down-conversion},
\newblock Phys. Rev. A \textbf{50}, 5122 (1994).

\bibitem{Aspelmeyer:Kippenberg:2014}
M.~Aspelmeyer, T.~J. Kippenberg and F.~Marquardt,
\newblock \emph{Cavity optomechanics},
\newblock Rev. Mod. Phys. \textbf{86}, 1391 (2014).

\bibitem{Walborn:Monken:2010}
S.~Walborn, C.~Monken, S.~Pádua and P.~S. Ribeiro,
\newblock \emph{Spatial correlations in parametric down-conversion},
\newblock Physics Reports \textbf{495}(4–5), 87  (2010).

\bibitem{Emary:Brandes:2003}
C.~Emary and T.~Brandes,
\newblock \emph{Chaos and the quantum phase transition in the dicke model},
\newblock Phys. Rev. E \textbf{67}, 066203 (2003).

\bibitem{Birrell:Davies:1984}
N.~D. Birrell and P.~C.~W. Davies,
\newblock \emph{Quantum Field in Curved Space},
\newblock Cambridge University Press (1984).

\bibitem{Arnowitt:Deser:2008}
R.~Arnowitt, S.~Deser and C.~Misner,
\newblock \emph{Republication of: The dynamics of general relativity},
\newblock General Relativity and Gravitation \textbf{40}(9), 1997 (2008).

\bibitem{Xiang:Ashhab:2013}
Z.-L. Xiang, S.~Ashhab, J.~Q. You and F.~Nori,
\newblock \emph{Hybrid quantum circuits: Superconducting circuits interacting
  with other quantum systems},
\newblock Rev. Mod. Phys. \textbf{85}, 623 (2013).

\bibitem{Wilson:Johansson:2011}
C.~M. Wilson, G.~Johansson, A.~Pourkabirian, M.~Simoen, J.~R. Johansson,
  T.~Duty, F.~Nori and P.~Delsing,
\newblock \emph{Observation of the dynamical casimir effect in a
  superconducting circuit},
\newblock Nature \textbf{479}(7373), 376 (2011).

\bibitem{Dalfovo:Giorgini:1999}
F.~Dalfovo, S.~Giorgini, L.~P. Pitaevskii and S.~Stringari,
\newblock \emph{Theory of bose-einstein condensation in trapped gases},
\newblock Rev. Mod. Phys. \textbf{71}, 463 (1999).

\bibitem{Carusotto:Ciuti:2013}
I.~Carusotto and C.~Ciuti,
\newblock \emph{Quantum fluids of light},
\newblock Rev. Mod. Phys. \textbf{85}, 299 (2013).

\bibitem{Hawking:1974}
S.~W. Hawking,
\newblock \emph{Black hole explosions?},
\newblock Nature \textbf{248}(5443), 30 (1974).

\bibitem{Alsing:Fuentes:2012}
P.~M. Alsing and I.~Fuentes,
\newblock \emph{Observer-dependent entanglement},
\newblock Classical and Quantum Gravity \textbf{29}(22), 224001 (2012).

\bibitem{Brown:MartinMartinez:2013}
E.~G. Brown, E.~Mart\'{i}n-Mart\'{i}nez, N.~C. Menicucci and R.~B. Mann,
\newblock \emph{Detectors for probing relativistic quantum physics beyond
  perturbation theory},
\newblock Phys. Rev. D \textbf{87}, 084062 (2013).

\bibitem{Mandel:Wolf:1995}
L.~Mandel and E.~Wolf,
\newblock \emph{Optical Coherence and Quantum Optics},
\newblock Cambridge University Press (1994).

\bibitem{Adesso:Ragy:2014}
G.~Adesso, S.~Ragy and A.~R. Lee,
\newblock \emph{Continuous variable quantum information: Gaussian states and
  beyond},
\newblock Open Systems \& amp; Information Dynamics \textbf{21}(01n02), 1440001
  (2014).

\bibitem{Williamson:1936}
J.~Williamson,
\newblock Am. J. Math. \textbf{58}, 141 (1936).

\bibitem{Blanes:Casas:2008}
S.~Blanes, F.~Casas, J.~Oteo and J.~Ros,
\newblock \emph{The magnus expansion and some of its applications},
\newblock Physics Reports \textbf{470}(5), 151  (2009).

\bibitem{NIST:2010}
R.~Roy and et.al.,
\newblock \emph{NIST Handbook of Mathematical Functions},
\newblock Cambridge University Press (2010).

\bibitem{Messiah:1961}
A.~Messiah,
\newblock \emph{Quantum Mechanics},
\newblock Dover Publications (1961).

\bibitem{Berman:James:2001}
G.~P. Berman, D.~F.~V. James and D.~I. Kamenev,
\newblock \emph{Stability of the ground state of a harmonic oscillator in a
  monochromatic wave},
\newblock Chaos \textbf{11}(3), 449 (2001).

\end{thebibliography}

\nolinenumbers

\end{document}